\documentclass[apj,appendixfloats]{emulateapj}
\usepackage{amsmath}
\usepackage{natbib}
\usepackage{graphicx}
\usepackage{appendix}
\usepackage{color}
\usepackage{relsize}
\usepackage{float}
\usepackage{enumitem}
\DeclareGraphicsExtensions{.eps}
\begin{document}
\title{Search for water in a super-Earth atmosphere: High-resolution optical spectroscopy of 55 Cancri~e}
\author{Lisa J. Esteves\altaffilmark{1}, 
Ernst J. W. De Mooij\altaffilmark{2,3}, 
Ray Jayawardhana\altaffilmark{4},
Chris Watson\altaffilmark{2} \& 
Remco de Kok\altaffilmark{5} }
\altaffiltext{1}{Astronomy \& Astrophysics, University of Toronto, 50 St. George Street, Toronto, Ontario M5S 3H4, Canada} 
\altaffiltext{2}{Astrophysics Research Centre, School of Mathematics and Physics, Queens University, Belfast, UK}
\altaffiltext{3}{School of Physical Sciences, Dublin City University, Dublin 9, Ireland}
\altaffiltext{4}{Physics \& Astronomy, York University, Toronto, Ontario L3T 3R1, Canada}
\altaffiltext{5}{Leiden Observatory, Leiden University, Postbus 9513, 2300 RA, Leiden, The Netherlands; SRON Netherlands Institute for Space Research, Sorbonnelaan 2, 3584 CA Utrecht, The Netherlands}
\begin{abstract}
\indent We present the analysis of high-resolution optical spectra of four transits of 55Cnc~e, a low-density, super-Earth that orbits a nearby Sun-like star in under 18 hours. The inferred bulk density of the planet implies a substantial envelope, which, according to mass-radius relationships, could be either a low-mass extended or a high-mass compact atmosphere. Our observations investigate the latter scenario, with water as the dominant species. We take advantage of the Doppler cross-correlation technique, high-spectral resolution and the large wavelength coverage of our observations to search for the signature of thousands of optical water absorption lines. Using our observations with HDS on the Subaru telescope and ESPaDOnS on the Canada-France-Hawaii Telescope, we are able to place a 3$\sigma$ lower limit of 10 g/mol on the mean-molecular weight of 55Cnc~e's water-rich (volume mixing ratio $>$10$\%$), optically-thin atmosphere, which corresponds to an atmospheric scale-height of $\sim$80 km. Our study marks the first high-spectral resolution search for water in a super-Earth atmosphere and demonstrates that it is possible to recover known water-vapour absorption signals, in a nearby super-Earth atmosphere, using high-resolution transit spectroscopy with current ground-based instruments.
\end{abstract}
\maketitle
\section{Introduction}\label{sec:intro}
\indent In recent years, enormous progress has been made in our understanding of the atmospheric properties of hot Jupiters through observations of transiting extrasolar planets. Notable discoveries include the first detection of sodium~\citep{Charbonneau2002}, the first detections of a secondary eclipse from space~\citep{Charbonneau2005, Deming2005} and the ground~\citep{deMooij2009, Sing2009} to detailed studies at high spectral resolution of carbon monoxide in day-side emission spectra~\citep{Brogi2012, Rodler2012}. Apart from the first detections, progress in our understanding of the atmospheric properties of hot Jupiters has been made through the analyses of transmission and emission spectra~\citep[e.g.][]{Swain2009,Deming2013,Stevenson2014, Sing2016}. For super-Earths, on the other hand, the available information on their atmospheric properties is very limited, mainly due to the much higher signal-to-noise ratio required to measure their atmospheric features, as well as the limited number of super-Earths known to orbit bright stars. \\
\indent One of the most studied super-Earths is GJ1214b, a very low density ($\rho$ $\sim$0.4 $\rho_{Earth}$) planet orbiting a M5V star. The planet is expected to either have an extended hydrogen-rich envelope or be water-rich and have a steam-atmosphere~\citep{Charbonneau2009}. The current observations \citep[e.g.][]{Bean2010, Bean2011, Kreidberg2014} point towards a featureless transmission spectrum, consistent with a thick cloud layer. HD 97658b, another well studied super-Earth~\citep{Knutson2014}, exhibits a featureless transmission spectrum, which suggests an atmosphere covered by thick clouds or composed of a molecular species much heavier than hydrogen. The observations of these two super-Earths were obtained with the 2.4 m Hubble Space Telescope, which has the advantage of being above the Earth's atmosphere but is limited by observing efficiency, especially for bright targets, and, due to the instruments available, also has limited wavelength coverage and spectral resolution. The largest ground-based telescopes, on the other hand, have much greater collecting areas than Hubble, and also host sophisticated new instruments, including high-resolution spectrographs capable of resolving individual absorption lines of molecules such as water. While the presence of the Earth's atmosphere has a significant negative impact on ground-based optical observations, the fact that the radial velocity of 55Cnc~e, its host star and Earth's atmosphere change very differently throughout the night allows for the use of cross-correlation with model spectra to distinguish the planetary signal from tellurics and stellar activity. \\
\begin{table*}[t]
\centering
\caption{Summary of Observations}
\setlength{\tabcolsep}{2pt}
\begin{tabular}{ccccccc}
Night & Date & Instrument/Telescope & Duration & $\#$ of & Cadence & Average SNR  \\
      & (UT) &                      & (hrs)    & Frames  & (s)     & of Continuum \\
\hline
\hline
1 & Feb. 9, 2014 & ESPaDOnS/CFHT & 4 & 76 & 187 & 150 \\
2 & Apr. 23, 2014 & ESPaDOnS/CFHT & 4 & 76 & 187 & 140 \\
3 & Dec. 12, 2014 & HDS/Subaru & 6 & 136 & 192 & 370 \\
4 & Jan. 9, 2015 & HDS/Subaru & 8.5 & 158 & 192 & 440 \\
\hline
\end{tabular}
\label{tab:obssum}
\end{table*}
\subsection{55Cnc~e}
\indent Of the known super-Earths, 55Cnc~e is the best target for atmospheric studies from the ground. It orbits a nearby G8V star, currently one of the brightest transiting planet hosts, in under 18 hours, making it one of the hottest planets known to date. The observations analyzed in this study take advantage of 55Cnc's brightness, and the large collecting area available from the ground, to attain the precision needed to detect a water-rich, low-mass, optically-thin atmosphere on 55Cnc~e. \\
\indent 55Cnc~e was initially discovered and characterized by radial-velocity measurements~\citep{McArthur2004, Fischer2008}. \citet{Dawson2010} later revised the analysis, yielding a period of 0.736539$\pm$0.000003 days and a minimum mass of 8.3$\pm$0.3 M$_{\oplus}$. Transits were subsequently detected by \citet{Winn2011} using the MOST satellite, and \citet{Demory2011} using the {\it Spitzer} Space Telescope. \citet{Winn2011} find 55Cnc~e's mass, radius and mean density to be 8.63$\pm$0.35 M$_{\oplus}$, 2.00$\pm$0.14 R$_{\oplus}$, and 5.9$\pm$1.5 g cm$^{-3}$, respectively, while \citet{Demory2011} find values of 7.8$\pm$0.5 M$_{\oplus}$, 2.1$\pm$0.2 R$_{\oplus}$ and 4.8$\pm$1.3 g cm$^{-3}$, respectively. Follow-up transit observations made from space by \citet{Gillon2012}, \citet{Dragomir2014} and \citet{Demory2016b} and from the ground by \citet{deMooij2014} report planet radius measurements in agreement with these studies. Altogether their results suggest that, since 55Cnc~e's mean density is comparable to that of Earth, despite the greater mass and consequently greater compression, the solid interior must be supplemented by a significant mass of water, gas, or other light elements. \\
\indent Although a purely silicate interior with no envelope was ruled out~\citep{Winn2011,Demory2011,Gillon2012}, \citet{Madhusudhan2012} suggest that a carbon-rich interior with no envelope could explain 55Cnc~e's inflated radius. However, follow-up secondary eclipse, phase-curve and transit measurements~\citep{Demory2012,Tsiaras2016,RiddenHarper2016,Demory2016a,Demory2016b} in conjunction with theoretical models~\citep{Lammer2013,Ito2015,Kite2016} point to the existence of an envelope. \\
\indent Mass-radius relationships for planets with envelopes composed of either hydrogen-helium or water were presented in \citet{Winn2011}, \citet{Demory2011} and \citet{Gillon2012}. These studies show that 55Cnc~e's density can be explained by either a low-mass hydrogen-helium or a high-mass water dominated atmosphere, but that simple atmospheric escape calculations show that the former is unlikely due to the short evaporation timescale, of order a million years, for hydrogen~\citep{Valencia2010}. Assuming the evaporation timescale of water is much longer, of order a billion years, they conclude that a water dominated envelope is more likely. However, \citet{Lammer2013} calculate the theoretical atmospheric mass-loss of a hydrogen-dominated envelope on 55Cnc~e that is being irradiated by XUV photons~\citep{Ehrenreich2012} and find that thermal mass-loss rates are $\sim$10 times lower than that of a typical hot-Jupiter, such as HD 209458b. As a result \citet{Lammer2013} conclude that one can expect that 55Cnc~e would not lose its hydrogen envelope during its remaining lifetime. \\
\indent \citet{Demory2012} report a {\it Spitzer} secondary eclipse depth of 131$\pm$28 ppm corresponding to a brightness temperature of 2360$\pm$300 K. \citet{Ito2015} find that this brightness temperature is in agreement with the theoretically predicted eclipse depth created by a mineral atmosphere on 55Cnc e. Their study investigates the radiative properties of the atmosphere that is in gas/melt equilibrium with the underlying magma ocean. Their equilibrium calculations yielded a mineral atmosphere (i.e. Na, K, Fe, Si, SiO, O, and O$_2$ as the major atmospheric species) and radiative absorption line data was used to calculate theoretical eclipse depths. \\
\indent \citet{Demory2016a, Demory2016b} observed 55Cnc~e's orbital phase curve and eclipse using {\it Spitzer} and report highly asymmetric day-side thermal emission, a strong day-night temperature contrast and a thermal day-side temperature that varies by a factor of 3.7 over a period of two years. These studies hypothesize that phase curve properties are the result of either an optically-thick atmosphere with heat recirculation confined to the planetary day-side, or a planet devoid of atmosphere with low-viscosity magma flowing at the surface. However, \citet{Kite2016} find, using basic models, that heat transport by magma currents is insufficient to explain the antistellar-hemisphere temperature found by \citet{Demory2016a} and that therefore an atmosphere is indicated. \\
\indent The composition and properties of a likely gaseous envelope on 55Cnc~e has also been the focus of recent space-based studies. \citet{Tsiaras2016} analyze Hubble Space Telescope observations and report the detection of features at 1.42 and 1.54 $\mu$m. Using Bayesian spectral retrieval, they conclude that these features are likely the result of trace amounts of HCN, retained within a light-weight atmosphere, presumably composed of mostly hydrogen and helium. Using ground-based data, \citet{RiddenHarper2016} report a tentative detection of an exosphere via absorption from sodium (3$\sigma$) and singly ionized calcium (4$\sigma$, variable). \\
\indent Our study, similar to \citet{RiddenHarper2016}, uses ground-based high-resolution transmission spectra to search for absorption signatures of 55Cnc~e's atmosphere. However, we take advantage of the large wavelength coverage and high spectral resolution of our observations to search for the signature of thousands of optical absorption lines produced by water. We note that, although 55Cnc~e has been called a possible water-world, the presence of was water is not expected if the planet's C/O abundance is $>$1~\citep{Hu2014}. However, it is unclear if 55Cnc~e is expected to be carbon-rich as its host star has been reported as both carbon-rich~\citep[][C/O=1.12$\pm$0.09]{DelgadoMena2010} and carbon-poor~\citep[][C/O=0.78$\pm$0.08]{Teske2013}.
\subsection{High-resolution cross-correlation of spectra}
\begin{table*}[t]
\centering
\caption{Stellar and Planetary Parameters Used in Analysis}
\setlength{\tabcolsep}{2pt}
\begin{tabular}{lccl}
Parameter & Value & \ Reference \\
\hline
\hline
Systemic velocity (km~s$^{-1}$) & 27.58 & \citet{Nidever2002} \\ 
Orbital Period (days) & 0.736542(3) & \citet{Dragomir2014} \\
Ephemeris (JD) & 2455962.067(2) & \citet{Dragomir2014} \\
Semi-major axis (au) & 0.01545(3) & \citet{Dragomir2014} \\
R$_p$/R$_{\star}$ & 0.019(8) & \citet{Dragomir2014} \\
a/R$_{\star}$ & 3.52(4) & \citet{Dragomir2014} \\
\hline
\end{tabular}
\label{tab:values}
\end{table*}
\indent Here we present the results of our search for water in four high-resolution ground-based transit observations of 55Cnc~e. To detect water we use a Doppler cross-correlation technique that combines the signal from thousands of water lines and uses the difference in Doppler shift to disentangle 55Cnc~e's spectra from its host star's and the Earth's atmospheric spectra. \\
\indent This technique was pioneered by \citet{Snellen2010} to detect carbon monoxide in the atmosphere of HD209458~b during the planet's transit. It has also been successfully used to detect carbon monoxide in the day-side spectra of $\tau$Bootis~b~\citep{Brogi2012,Rodler2012}, as well as carbon monoxide and water in the transmission spectra of $\beta$Pic~b~\citep{Snellen2014}, HD189733~b~\citep{Brogi2016,Lockwood2014,Dekok2013,Birkby2013,Rodler2013b}, HD179949~b~\citep{Brogi2014} and 51Peg~b~\citep{Brogi2013}. In addition, a similar study in the optical has claimed to have detected reflected light from 51Peg~b~\citep{Martins2015} and a study at ultraviolet wavelengths claims a low-significance detection of reflected light from $\tau$Bootis~b~\citep{Rodler2013a}.\\
\indent We apply this technique to search for water at optical wavelengths in the transmission spectrum of the super-Earth 55Cnc~e. We present our high-resolution observations of four transits in Section~\ref{sec:obs}. In Section~\ref{sec:dr} we describe the data reduction, which includes extracting and normalizing the 1D spectra, then removing contamination from systematics and stellar and telluric absorption. In our analysis, in Section~\ref{sec:ana}, we describe our cross-correlation and retrieval procedure as well as the atmospheric models used. We present and discuss our results in Section~\ref{sec:res}.
\section{Observations}\label{sec:obs}
\indent We observed two transits of 55Cnc~e with the Echelle SpectroPolarimetric Device for the Observation of Stars~\citep[ESPaDOns;][]{espadons2003} at the Canada-France Hawai'i Telescope (CFHT). The first of these transits was observed on the night of February 9, 2014 UT (hereafter N$_1$) and the second on April 23, 2014 UT (herafter N$_2$). The observations were done in the Queued Service Observing mode, where the observations are executed by the staff at CFHT. For both nights we used the instrument in the `Star+Sky' mode, resulting in a resolution of R$\sim$68,000. We used an exposure time of 149 seconds, resulting in an average cadence of $\sim$187 seconds. On each night we observed for just over 4 hours and obtained 76 frames per night, of which 29 were during transit. The weather during N$_{1}$ was photometric throughout the observations, with a median seeing of 0.5$^{\prime\prime}$. During N$_{2}$ the weather was photometric at the start, with some thin cirrus towards the end of the observations, and seeing varied between $\sim$0.7$^{\prime\prime}$ and $\sim$1.1$^{\prime\prime}$. The average continuum SNR of the spectra were approximately 150 and 140 for  N$_{1}$ and  N$_{2}$, respectively. The extracted CFHT spectra cover a wavelength range of 5060-7950 $\AA$ spanning 16 orders. \\
\indent We also observed two transits with the High Dispersion Spectrograph~\citep[HDS;][]{Noguchi2002} at the Subaru Telescope. We observed one transit on December 12th, 2014 (hereafter N$_3$) and a second on January 9th, 2015 (hereafter N$_4$). For both nights the observations were taken using the \#1 image slicer rather than the slit to allow for a high signal-to-noise ratio at a spectral resolution of R$\sim$110,000~\citep{Tajitsu2012}. To reduce the overheads, we binned by two pixels in the spatial direction, leaving the binning in the spectral direction unchanged. We set the exposure to 120 seconds, resulting in an average cadence of $\sim$192 seconds. We observed the first transit for just over 6 hours and the second for 8.5 hours, during which we obtained 136 and 158 total frames. During each night 28 frames were taken during transit. The weather during N$_{3}$ was photometric throughout the observations, with a median seeing of 0.5$^{\prime\prime}$. During N$_{4}$ the weather was also photometric but the seeing varied between $\sim$0.4$^{\prime\prime}$ and $\sim$0.7$^{\prime\prime}$. The average continuum SNR of the spectra were approximately 370 and 440 for  N$_{3}$ and  N$_{4}$, respectively. The extracted Subaru spectra cover a wavelength range of 5240-7890 $\AA$ spanning 38 orders. A summary of the observations can be found in Table~\ref{tab:obssum}.
\section{Data reduction and correction of systematic effects}\label{sec:dr}
\begin{figure*}[t]
\begin{center}
%\scalebox{0.83}{\includegraphics{fig_corr/fig_4n_corr_remco_H2O_050.eps}}
\scalebox{0.83}{\includegraphics{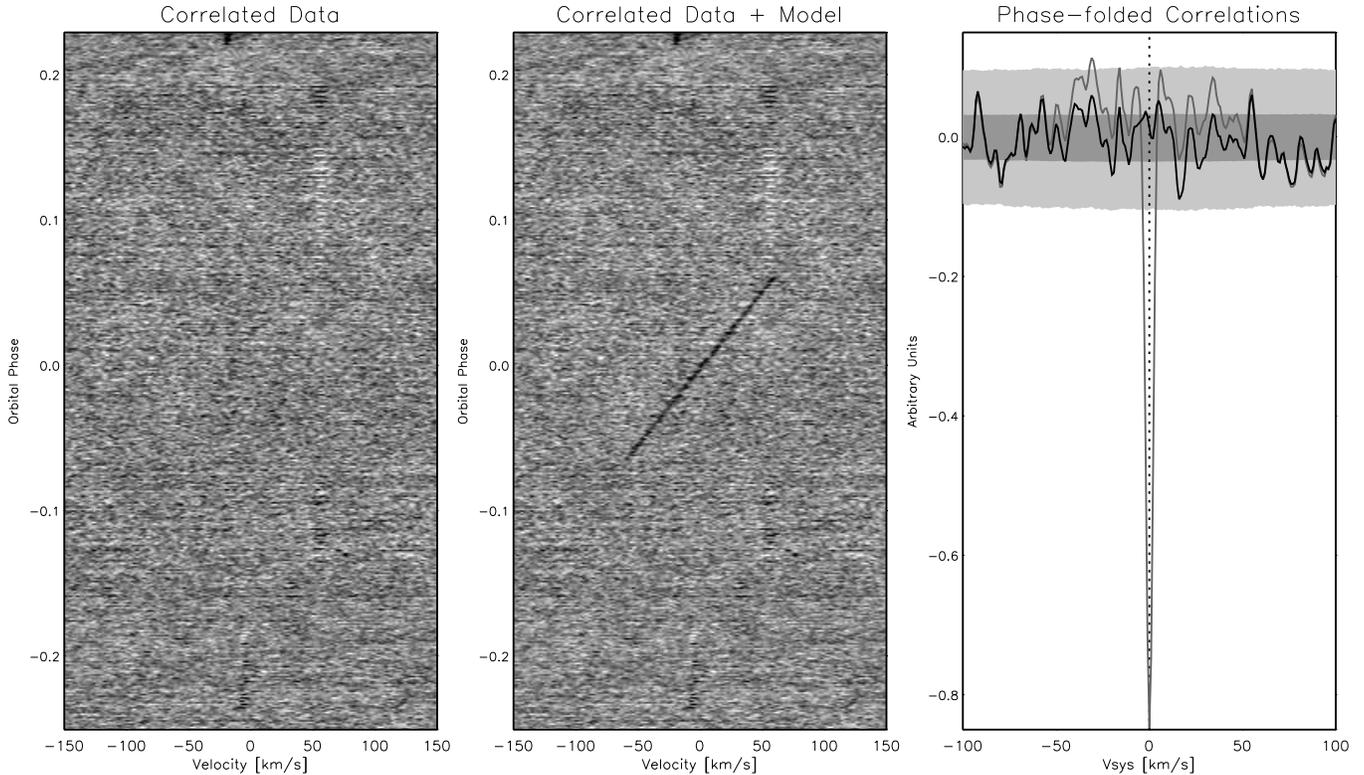}}
\end{center}
\caption{Left panel: Correlated data from all four nights of observations. Middle panel: Correlated data + model with a water VMR of 10\% and a low mean molecular weight of 2 g/mol. Right panel: Correlation phase-folded to the planet's orbital radial velocity. The black line is the data and the grey line is the data + model. The dark grey region represents 1$\sigma$ values for each data point while the light grey region represents the 3$\sigma$ values. For a description of how these were calculated see Section~\ref{sec:err}.}
\label{fig:corr}
\end{figure*}
\begin{figure*}[t]
\begin{center}
%\scalebox{0.83}{\includegraphics{fig_corr/fig_4n_SNR300a_corr_remco_H2O_050.eps}}
\scalebox{0.83}{\includegraphics{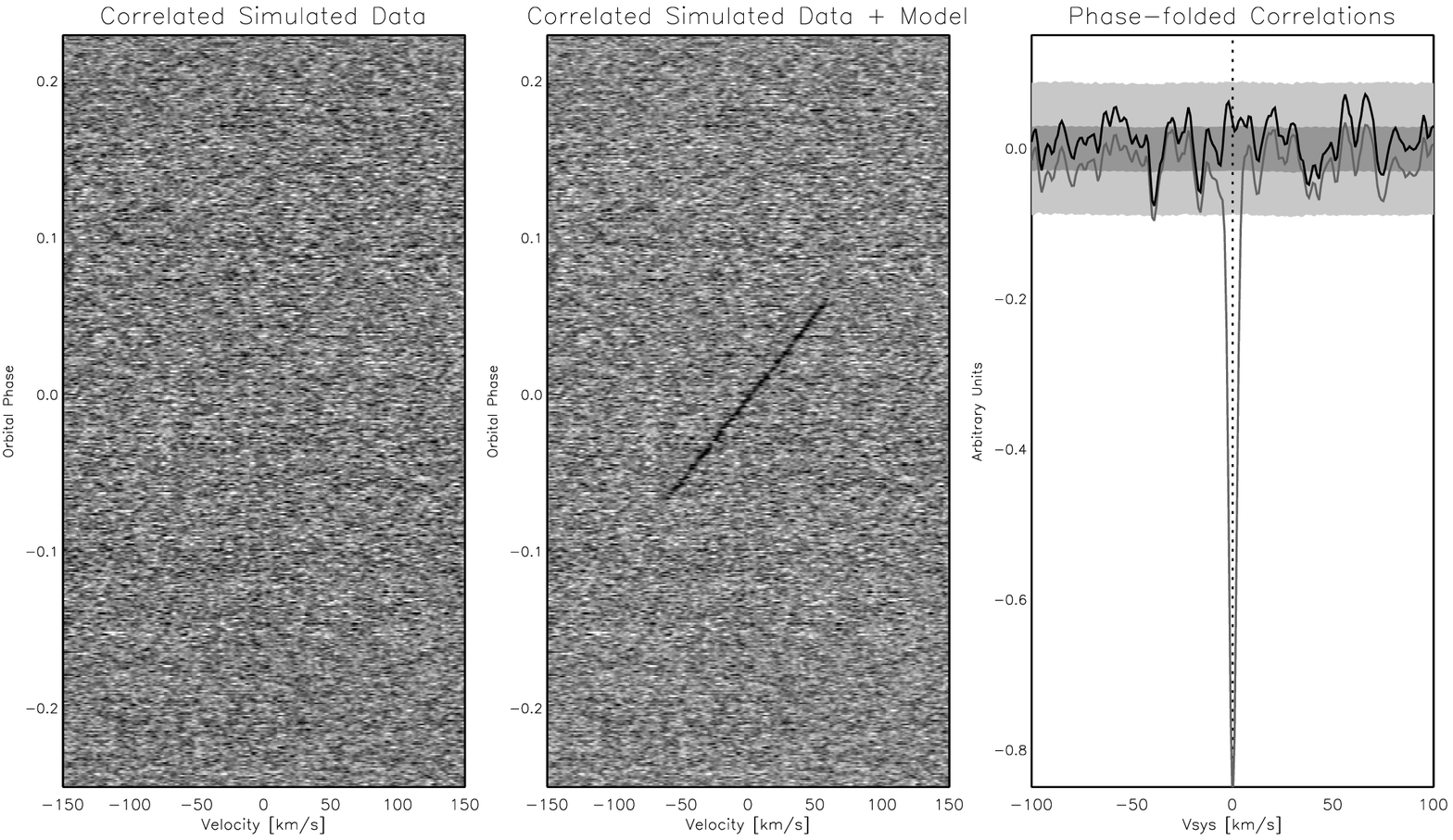}}
\end{center}
\caption{Correlated and phase-folded simulated data, with normally distributed noise with a signal-to-noise ratio of 300. For description of each panel see Fig.~\ref{fig:corr}.}
\label{fig:corr_sim}
\end{figure*}
\subsection{Extraction and alignment of the 1D spectra}\label{sec:dr1}
\indent The CFHT observations (N$_1$ and N$_2$) were reduced and extracted using the Upena pipeline at CFHT which is based on Libre ESpRIT~\citep{Donati1997}. We use the unnormalized but wavelength corrected spectra produced by the pipeline for the rest of our analysis. Although the pipeline corrects for some drifts in the wavelength solution, we determine any residual shifts directly. To do this we measure the position of strong telluric lines due to oxygen and water. For N$_1$ and N$_2$ the maximum velocity drift of the telluric lines was less than 0.2 km/s, which corresponds to a drift in wavelength of less than 0.0006 $\AA$. We subsequently interpolate all frames onto a common wavelength grid in the telluric reference frame taking these small drifts into account. The extracted CFHT spectra cover a wavelength range of 5060-7950 $\AA$ spanning 16 orders and the output from this stage, for three orders, can be seen in the top sub-panels of Appendix Figures~\ref{fig:reduct1}~and~\ref{fig:reduct2}.\\
\indent The Subaru observations (N$_3$ and N$_4$) were reduced using a combination of IRAF scripts and custom IDL routines. As a first step the overscan correction was performed for all frames using the {\it overscan.cl} IRAF script\footnote{http://www.subarutelescope.org/Observing/Instruments/ HDS/hdsql/overscan.cl} provided by the observatory. Subsequently the detector non-linearity~\citep{Tajitsu2012} was corrected using a script provided on the instrument pages\footnote{http://www.subarutelescope.org/Observing/Instruments/ HDS/hdsql/hdslinear.cl}. The individual bias frames were combined in IRAF and subtracted from all remaining frames. \\
\indent When inspecting the bias-subtracted frames, we identified and removed a low-level crosstalk, at the level of 1.4$\cdot$10$^{-3}$, between the two halves of each of the CCDs. The level was determined by varying the crosstalk coefficient until the signal was no longer detectable. \\
\indent We subsequently combined the individual bias-subtracted flat frames into an uncorrected master flat field using IRAF, and used this to trace the echelle orders. To trace the location of the echelle orders we determined, at each wavelength, the position of the points on the edge of the echelle order where the flux was 10\% of the peak flux at that wavelength. We then fit a 2D polynomial to the position of all the echelle orders on an individual CCD, in order to retrieve a global solution. This solution was used for both order masking and spectral extraction. \\
\indent Subsequently, we removed the scattered light from the master-flat. The scattered light was fit on a wavelength-by-wavelength basis by fitting a low-order polynomial to the light between orders for each CCD along the spatial axis, after masking out the orders based on the trace performed earlier. We generated our final, normalized master flat by fitting the flux in the individual orders as a function of wavelength and dividing the unnormalised flat by this fit.\\
\indent The science frames were subsequently flat fielded, and we used custom IDL scripts to perform the trace of the spatial centroid of the star in each order, as a function of wavelength. We performed the trace by simultaneously fitting Gaussian profiles to each of the image slices, while fixing the distance between the stellar psf in each of the slices. During the fit, we used a common FWHM for all the slices, but the amplitude for each slice was allowed to vary. As with the flatfield, a 2d polynomial was fit to the individual traces in order to obtain a global solution for each CCD. \\
\indent This trace was used to help identify and correct cosmic rays and bad pixels using a custom IDL script. We determined the average profile in segments with a length of 500 pixels in the spectral direction on an order-by-order basis. These segments were generated in steps of 250 pixels, and the average profile corresponding to the nearest segment to the wavelength under consideration was used. This average profile was scaled to match the local profile, and any pixel discrepant by more than 5$\sigma$ was replaced by the scaled average profile. \\
\indent After correction for cosmic rays, we removed scattered light in the same way as for the master flat. Finally we extracted the spectra by summing up the flux over the full spatial extent of each order, thereby directly combining the flux from all five image slices. \\
\indent Wavelength calibrations were made using ThAr frames taken just before and after the observations of 55Cnc. The positions of the ThAr lines were identified and fitted using the {\it ecidentify} routine in IRAF. In order to maximize observing efficiency we did not take any ThAr frames during the science observations, but instead relied on the telluric absorption features to track drifts in the wavelength solution. For N$_3$ and N$_4$ the maximum velocity drift of the telluric lines was less than 0.4 km/s, which corresponds to a drift in wavelength of less than 0.002 $\AA$. We finally interpolated all spectra within a night onto a common wavelength grid in the telluric reference frame. The extracted Subaru spectra cover a wavelength range of 5240-7890 $\AA$ spanning 38 orders and the output from this stage, for three orders, can be seen in the top sub-panels of Appendix Figures~\ref{fig:reduct3}~and~\ref{fig:reduct4}.
\subsection{Normalization of spectra}\label{sec:dr2}
\indent Large scale time-dependent systematics, whose likely cause is the instrument's changing blaze response, were removed, and the spectra were simultaneously normalized. To do this we first divided an individual spectrum on a night by a nightly reference image, for which we used the first frame of the night. The divided spectrum was then binned by 201 pixels ($\sim$5 $\AA$) for the CFHT data and 41 pixels ($\sim$1 $\AA$) for the Subaru data, fit with a spline and evaluated at the wavelength positions of the unbinned spectrum. The evaluated spline was then divided out of the original spectrum, resulting in the removal of the time-variable blaze response and the normalization of the each spectrum continuum to the reference frame continuum. We then removed the 100 pixels at the edges of each order in order to avoid contamination from the poor constraints at the edge. We note that, due to the shape of the blaze function, these pixels already have a significantly lower SNR compared to the centre of each order, and therefore removing these pixels has minimal impact on the results. \\
\indent We selected these binsizes in order to avoid removing the absorption lines from the planetary atmosphere, while allowing us to correct the small wavelength scale variations seen in the blaze function for the Subaru observations. This processing results in frames from a single night sharing the same overall blaze function, while maintaining their individual high-frequency signal. The results from the normalisation can be seen in the upper middle sub-panels of Appendix Figures~\ref{fig:reduct1}~to~\ref{fig:reduct4}.
\subsection{Removal of telluric and stellar lines}\label{sec:dr3}
\indent The stellar and telluric lines were removed using the SYSREM algorithm described by \citet{Tamuz2005}. Both in-transit and out-of-transit frames from an individual night were used to determine the correction. Each order was treated separately and for consistency we chose to apply six iterations of SYSREM to each night of data. We chose to apply the maximum recommended number of iterations because the RMS of the residuals for several orders does not plateau until six iterations have been applied. After SYSREM was applied the averaged RMS of each order for Night 1 to 4 was 0.004, 0.005, 0.002 and 0.001, respectively. \\
\indent This method is able to remove the telluric and stellar absorption features, that are stable in time, while preserving the signal from 55Cnc~e's atmosphere. With an average orbital velocity of 228.7 km~s$^{-1}$, 55Cnc~e's radial orbital velocity changes rapidly during the transit, from -57.6 to +57.6 km~s$^{-1}$. Our reduction does poorly in the orders with strong closely spaced telluric lines, in particular between 7600 and 7800 $\AA$ where many strong oxygen lines are present. The strong lines make properly modelling the blaze function difficult and as a result introduce non-linear variations that are not removed by SYSREM. Similar to \citet{Snellen2010}, before cross-correlating, and to reduce contamination from noisy frames, we weighted each of the frames by dividing by the frames' standard deviation. We then similarly weighted each pixel, by dividing by the standard deviation of the pixel's variations throughout the night. This was done to suppress the contribution from noisy frames pixels. The results for this step can be seen in the lower middle sub-panels of Appendix Figures~\ref{fig:reduct1}~to~\ref{fig:reduct4}, while a plot of pixel RMS for these orders is shown in the bottom sub-panel.
\section{Analysis}\label{sec:ana}
\indent Our analysis of the data relies on the cross-correlation of the spectra with absorption models. This method requires very high-resolution data, since the precision of the cross-correlation relies on the ability to resolve individual lines. The precision also increases with the number of absorption lines and therefore molecules, whose rotation-vibration transitions produce thousands of absorption lines, are good targets for this type of analysis. 
\subsection{Atmospheric models}
\indent In order to search for the Doppler shifted signal from the thousands of absorption lines of water, we used a model that was calculated specifically for 55Cnc~e using molecular data from the high-temperature molecular spectroscopic database~\citep[HITEMP;][]{Rothman2010}, assuming Voigt line profiles, with a line wing cutoff of 50 cm$^{-1}$, and temperature broadened using the standard database parameters. Furthermore, Rayleigh scattering was included. Opacities were calculated line-by-line on a grid of 50 atmospheric layers from 5 down to 10$^{-10}$ bar and assuming a uniform temperature of 1000 K for pressures below 0.01 bar, 1500 K for pressures from 0.01 to 1.0 bar and 2500 K for pressures about 1.0 bar. This temperature-pressure profile agrees with the profile used by \citet{Demory2016a}. However, we note that the spectral shape is not a strong function of temperature. Parameters used in these calculations are the stellar radius (0.943 R$_{\odot}$), planetary radius (2.1 R$_{\oplus}$) and surface gravity at the bottom atmospheric layer (17.3 m$\cdot$s$^{-2}$ derived using a planet mass of 7.8 M$_{\oplus}$). Opacities were integrated along slant paths from the direction of the star to the observer. Integration over altitude and location on the disc then yielded the transmission of the planet for a given wavelength. \\
\indent Models were calculated for a grid of atmospheric volume mixing ratios (VMR = 10\%, 1\%, 0.1\% and 0.01\%) and mean molecular weights ($\mu$ = 2, 5, 10, 15 g/mol) over a wavelength range of 5000-8000 $\AA$ with 60,000 evenly spaced data points. See Appendix Fig.~\ref{fig:model} for an example plot of one of the models used in this analysis. In general, using a lower volume mixing ratio significantly reduces the strength of absorption features at the blue end of our spectral data, due to masking by Rayleigh scattering, while increasing the weight of the atmosphere scales down the strength of the lines equally for all wavelengths. Before use, all models were convolved to the resolution of the data they were used with. 
\subsection{Cross-correlation}
\indent For each night, each spectrum was correlated with models at Doppler shifts spanning -250 km~s$^{-1}$ to +250 km~s$^{-1}$ in steps of 1 km~s$^{-1}$. We then phase fold the correlation signal from the in-transit frames. To do this we first shift each correlation to the reference frame of 55Cnc~e, in order to correct for the Earth's rotation and orbital motion, the star's systemic velocity and the planet's radial velocity (see Table~\ref{tab:values}). The correlations are then interpolated to a common velocity grid using a cubic spline and in-transit frames were selected using transit models calculated via \citet{Mandel2002} and parameters in Table~\ref{tab:values}. Each point in velocity space is then summed, resulting in a measurement of model correlation strength as a function of systemic velocity for the in-transit data, where we expect to see the signal from the planet at V$_{sys}$=0 km~s$^{-1}$.
\subsection{Model recovery tests and estimation of the detection significance}\label{sec:err}
\indent In order to assess our ability to detect and constrain the properties of the atmosphere of 55Cnc~e, we ran several injection/recovery tests. To inject a model we multiplied each in-transit spectrum by an atmospheric model in the exoplanet's reference frame. This was done after aligning the spectra onto a common wavelength grid, but before any further steps. The data with the models injected were processed and analysed in an identical way to the normal science data. By injecting a water signal of varying strength into the data, and determining which can be recovered by our analysis, we are able to place sensitivity limits on our results. \\
\indent Assessing the detection significance is important. We therefore randomly drew 28 corrected spectra from all the corrected spectra during a night, assigned an in-transit phase to each of the spectra, and subsequently correlated and phase folded the selected spectra. We repeated this procedure 10,000 times in order to assertain the 1$\sigma$ and 3$\sigma$ confidence intervals. To determine the significance of our detections we compare these confidence intervals to the difference in correlation strength at V$_{sys}$=0 km~s$^{-1}$ of the normal and injected data. \\
\indent We also reran our injection/recovery test to assess the affect of interpolating the data to a common wavelength grid after injecting the model. In these tests we use the same procedure, but instead inject the water model (with a mean molecular mass of 10 g/mol and a water VMR of 10\%) before interpolating to a common wavelength grid. A correlation of the data with the model injected before and after is shown in Appendix Figure~\ref{fig:inject_before}. The differences between these correlations are not significant in comparison to the noise, which gives us confidence that the wavelength interpolation is not affecting our final results.
\section{Results and Discussion}\label{sec:res}
\begin{figure*}[t!]
\begin{center}
%\scalebox{0.83}{\includegraphics{fig_results/fig_4n_results_remco_55CNCe_37.eps}}
\scalebox{0.83}{\includegraphics{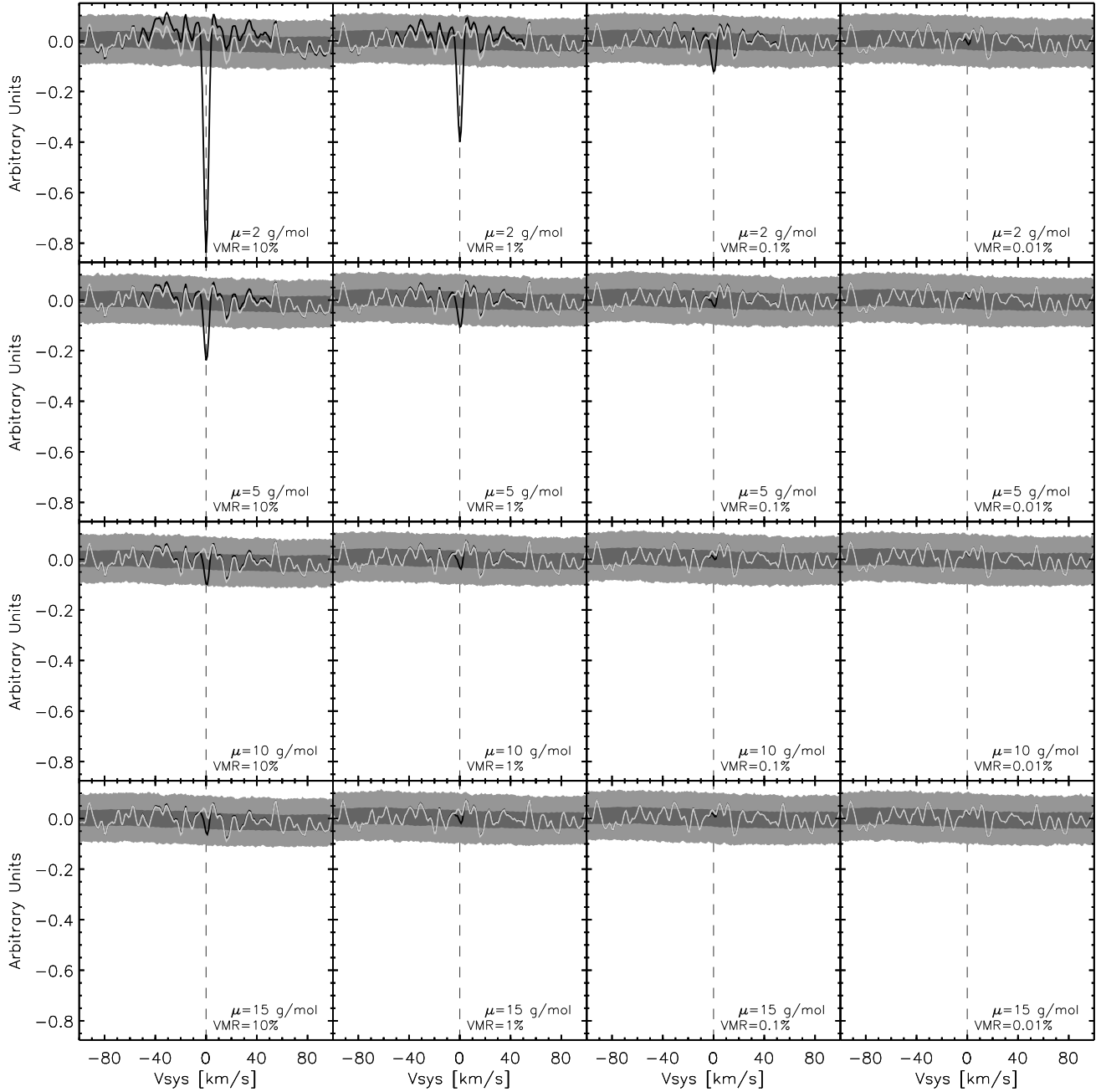}}
\end{center}
\caption{Phase folded correlations for a range of models that vary in volume mixing ratio and mean molecular mass. The black line is the data and the grey line is the data + model. The dark grey region represents 1$\sigma$ values for each data point while the light grey region represents the 3$\sigma$ values.}
\label{fig:results}
\end{figure*}
\indent The correlation of our original and injected data can be found in Fig.~\ref{fig:corr} along with a plot of the correlation after phase folding to 55Cnc~e's orbital velocity. The model injected in Fig.~\ref{fig:corr} utilizes a high water VMR of 10\% and a low mean molecular weight of 2 g/mol; it can easily be seen in the injected correlation as a dark diagonal trace in the in-transit frames. The slope of the trace corresponds to the planet's radial velocity change during the transit with respect to the systemic velocity. We again note that our reduction was not able to remove all contamination from tellurics. Residuals can be seen in the out-of-transit correlations in the left and centre panels of Fig.~\ref{fig:corr}. The features at phases 0.05 to 0.2 and $\sim$60 km~s$^{-1}$ are telluric residuals from Night 2, where the combined heliocentric and systemic velocity is  $\sim$60 km~s$^{-1}$. However, since the correlations used in our final analysis are only from the in-transit frames, this contamination should not influence our results. \\
\indent In the phase folded plot in the right panel of Fig.~\ref{fig:corr}, the injected signal is detected at $>$20$\sigma$ at V$_{sys}$=0. However, the original data exhibits no significant features in both the correlated and phase-folded data. \\
\indent Although there is no significant ($>$3$\sigma$) peak in our data, there are many peaks at more than 1$\sigma$. Since the many large features could hide the signal from the planet, we want to understand their cause. As the spectrum from water is very closely spaced, it is likely that noise spikes are picked up by adjacent water lines. To test this, we simulated a dataset with purely white noise at a signal-to-noise ratio of 300. We correlated and phase-folded these simulated spectra (both with and without a planetary signal injected). We also used the same analysis to assess the significance of any features. The results from this analysis can be seen in Fig.~\ref{fig:corr_sim}. It is clear that the white noise gives rise to a similar structure. \\
\indent Although Fig.~\ref{fig:corr} demonstrates that we can clearly rule out a large scale-height, very water-rich model, it does not allow us to immediately constrain the composition and scale-height of 55Cnc~e's atmosphere. We therefore inject models at different VMR and $\mu$ to see which model we would be able to detect at 3$\sigma$. The phase folded correlations are shown in Fig.~\ref{fig:results}. \\
\indent By comparing the phase folded correlations, normal and injected, we are able to place a 3$\sigma$ lower limit of 10 g/mol on the mean-molecular weight of 55Cnc~e's water-rich (volume mixing ratio $>$10$\%$), optically-thin atmosphere, which corresponds to an atmospheric scale-height of $\sim$80 km. This implies that if 55Cnc~e has a water-rich atmosphere, as suggested by \citet{Gillon2012} and \citet{Winn2011}, and is optically-thin, it must have a mean molecular weight heavier than 10 g/mol. The phase folded correlations of the water-depleted atmosphere models, 1$\%$ and 0.1$\%$ water by volume, yielded a 3$\sigma$ lower limit of 5 and 2 g/mol, respectively, while we could not detect a signal from a 0.01$\%$ water atmosphere. Our result, that 55Cnc~e does not have a water-rich, optically-thin extended atmosphere, is not surprising as water, with a mean molecular mass of 18 g/mol, would produce a heavy, and therefore compact, atmosphere if it were a large fraction by mass. \\
\indent We have shown that if 55Cnc~e has an optically-thin extended atmosphere, it is not water-rich. Evidence for the existence of an extended atmosphere on the planet includes detections reported by \citet{Tsiaras2016} of features at 1.42 and 1.54 $\mu$m, which they attribute to trace amounts of HCN retained within the planet's extended atmosphere. Additional evidence includes the tentative detection of sodium and singly ionized calcium reported by \citet{RiddenHarper2016}, as well as findings from \citet{Lammer2013} that 55Cnc~e's theoretical mass-loss rate is $\sim$10 times lower than that of a typical hot-Jupiter and that therefore 55Cnc~e can retain a hydrogen envelope. Furthermore, a water-depleted atmosphere was predicted by \citet{Ito2015}, who modelled an atmosphere that is in gas/melt equilibrium with an underlying magma ocean, a possible scenario for 55Cnc~e given its high temperature. They found that their equilibrium calculations yielded a water-depleted mineral atmosphere (i.e. Na, K, Fe, Si, SiO, O, and O$_2$ as the major atmospheric species). \\
\indent We note that, although the opacity of 55Cnc~e's atmosphere is not known, our mean molecular mass limits assume that the planet's atmosphere is optically-thin (i.e. the signal is not being masked by clouds or hazes). Indirect evidence that 55Cnc~e possibly harbours an optically-thick atmosphere was presented by \citet{Demory2016b, Demory2016a} and corroborated by \citet{Kite2016}. The latter study finds that heat transport by magma cannot explain the antistellar-hemisphere temperature found by \citet{Demory2016a} and concludes that an optically-thick atmosphere is likely required to transport heat to the planet's night-side. \\
\indent We conclude that our results are compatible with a cloudy atmosphere as well as an optically-thin atmosphere. In the cloudy scenario we cannot place any constraints on the planet's weight or composition, due to the fact that absorption features can be partially or fully obscured by clouds and other aerosols. In the optically-thin scenario our result is compatible with a light-weight atmosphere that is depleted of water (VMR$<$10\%) as well as a heavy atmosphere with any composition, including water-rich.  \\
\indent While our analysis did not detect water on 55Cnc~e, our injection and recovery tests demonstrate that it is possible to recover known water-vapour absorption signals, in a nearby super-Earth atmosphere, using high-resolution transit spectroscopy with current ground-based instruments. At the moment, only a couple of nearby super-Earth are known. However, upcoming surveys and missions, such as TESS, are expected to reveal many terrestrial planets, some of which will be excellent targets for follow-up observations of their atmospheres using the high-resolution Doppler cross-correlation technique.
\acknowledgments
This work was based in part on data collected on HDS at the Subaru Telescope, which is operated by the National Astronomical Observatory of Japan and on data collected by ESPaDOnS at the Canada-Hawaii-France Telescope, which is operated by the National Research Council (NRC) of Canada, the Institut National des Science de l'Univers of the Centre National de la Recherche Scientifique (CNRS) of France, and the University of Hawaii. This work was supported by grants to R.J. from the Natural Sciences and Engineering Research Council (NSERC) of Canada. L.J.E is supported in part by an NSERC CGS while EdM was funded by the Michael West Fellowship and C.A.W. acknowledges support by STFC grant ST/L000709/1.
\newpage
%\bibliography{references}

\begin{thebibliography}{57}
\expandafter\ifx\csname natexlab\endcsname\relax\def\natexlab#1{#1}\fi

\bibitem[{{Bean} {et~al.}(2010){Bean}, {Miller-Ricci Kempton}, \&
  {Homeier}}]{Bean2010}
{Bean}, J.~L., {Miller-Ricci Kempton}, E., \& {Homeier}, D. 2010, \nat, 468,
  669

\bibitem[{{Bean} {et~al.}(2011){Bean}, {D{\'e}sert}, {Kabath}, {Stalder},
  {Seager}, {Miller-Ricci Kempton}, {Berta}, {Homeier}, {Walsh}, \&
  {Seifahrt}}]{Bean2011}
{Bean}, J.~L., {D{\'e}sert}, J.-M., {Kabath}, P., {et~al.} 2011, \apj, 743, 92

\bibitem[{{Birkby} {et~al.}(2013){Birkby}, {de Kok}, {Brogi}, {de Mooij},
  {Schwarz}, {Albrecht}, \& {Snellen}}]{Birkby2013}
{Birkby}, J.~L., {de Kok}, R.~J., {Brogi}, M., {et~al.} 2013, \mnras, 436, L35

\bibitem[{{Brogi} {et~al.}(2016){Brogi}, {de Kok}, {Albrecht}, {Snellen},
  {Birkby}, \& {Schwarz}}]{Brogi2016}
{Brogi}, M., {de Kok}, R.~J., {Albrecht}, S., {et~al.} 2016, \apj, 817, 106

\bibitem[{{Brogi} {et~al.}(2014){Brogi}, {de Kok}, {Birkby}, {Schwarz}, \&
  {Snellen}}]{Brogi2014}
{Brogi}, M., {de Kok}, R.~J., {Birkby}, J.~L., {Schwarz}, H., \& {Snellen},
  I.~A.~G. 2014, \aap, 565, A124

\bibitem[{{Brogi} {et~al.}(2012){Brogi}, {Snellen}, {de Kok}, {Albrecht},
  {Birkby}, \& {de Mooij}}]{Brogi2012}
{Brogi}, M., {Snellen}, I.~A.~G., {de Kok}, R.~J., {et~al.} 2012, \nat, 486,
  502

\bibitem[{{Brogi} {et~al.}(2013){Brogi}, {Snellen}, {de Kok}, {Albrecht},
  {Birkby}, \& {de Mooij}}]{Brogi2013}
---. 2013, \apj, 767, 27

\bibitem[{{Charbonneau} {et~al.}(2002){Charbonneau}, {Brown}, {Noyes}, \&
  {Gilliland}}]{Charbonneau2002}
{Charbonneau}, D., {Brown}, T.~M., {Noyes}, R.~W., \& {Gilliland}, R.~L. 2002,
  \apj, 568, 377

\bibitem[{{Charbonneau} {et~al.}(2005){Charbonneau}, {Allen}, {Megeath},
  {Torres}, {Alonso}, {Brown}, {Gilliland}, {Latham}, {Mandushev}, {O'Donovan},
  \& {Sozzetti}}]{Charbonneau2005}
{Charbonneau}, D., {Allen}, L.~E., {Megeath}, S.~T., {et~al.} 2005, \apj, 626,
  523

\bibitem[{{Charbonneau} {et~al.}(2009){Charbonneau}, {Berta}, {Irwin}, {Burke},
  {Nutzman}, {Buchhave}, {Lovis}, {Bonfils}, {Latham}, {Udry}, {Murray-Clay},
  {Holman}, {Falco}, {Winn}, {Queloz}, {Pepe}, {Mayor}, {Delfosse}, \&
  {Forveille}}]{Charbonneau2009}
{Charbonneau}, D., {Berta}, Z.~K., {Irwin}, J., {et~al.} 2009, \nat, 462, 891

\bibitem[{{Dawson} \& {Fabrycky}(2010)}]{Dawson2010}
{Dawson}, R.~I., \& {Fabrycky}, D.~C. 2010, \apj, 722, 937

\bibitem[{{de Kok} {et~al.}(2013){de Kok}, {Brogi}, {Snellen}, {Birkby},
  {Albrecht}, \& {de Mooij}}]{Dekok2013}
{de Kok}, R.~J., {Brogi}, M., {Snellen}, I.~A.~G., {et~al.} 2013, \aap, 554,
  A82

\bibitem[{{de Mooij} {et~al.}(2014){de Mooij}, {L{\'o}pez-Morales},
  {Karjalainen}, {Hrudkova}, \& {Jayawardhana}}]{deMooij2014}
{de Mooij}, E.~J.~W., {L{\'o}pez-Morales}, M., {Karjalainen}, R., {Hrudkova},
  M., \& {Jayawardhana}, R. 2014, \apjl, 797, L21

\bibitem[{{de Mooij} \& {Snellen}(2009)}]{deMooij2009}
{de Mooij}, E.~J.~W., \& {Snellen}, I.~A.~G. 2009, \aap, 493, L35

\bibitem[{{Delgado Mena} {et~al.}(2010){Delgado Mena}, {Israelian},
  {Gonz{\'a}lez Hern{\'a}ndez}, {Bond}, {Santos}, {Udry}, \&
  {Mayor}}]{DelgadoMena2010}
{Delgado Mena}, E., {Israelian}, G., {Gonz{\'a}lez Hern{\'a}ndez}, J.~I.,
  {et~al.} 2010, \apj, 725, 2349

\bibitem[{{Deming} {et~al.}(2005){Deming}, {Seager}, {Richardson}, \&
  {Harrington}}]{Deming2005}
{Deming}, D., {Seager}, S., {Richardson}, L.~J., \& {Harrington}, J. 2005,
  \nat, 434, 740

\bibitem[{{Deming} {et~al.}(2013){Deming}, {Wilkins}, {McCullough}, {Burrows},
  {Fortney}, {Agol}, {Dobbs-Dixon}, {Madhusudhan}, {Crouzet}, {Desert},
  {Gilliland}, {Haynes}, {Knutson}, {Line}, {Magic}, {Mandell}, {Ranjan},
  {Charbonneau}, {Clampin}, {Seager}, \& {Showman}}]{Deming2013}
{Deming}, D., {Wilkins}, A., {McCullough}, P., {et~al.} 2013, \apj, 774, 95

\bibitem[{{Demory} {et~al.}(2016{\natexlab{a}}){Demory}, {Gillon},
  {Madhusudhan}, \& {Queloz}}]{Demory2016a}
{Demory}, B.-O., {Gillon}, M., {Madhusudhan}, N., \& {Queloz}, D.
  2016{\natexlab{a}}, \mnras, 455, 2018

\bibitem[{{Demory} {et~al.}(2012){Demory}, {Gillon}, {Seager}, {Benneke},
  {Deming}, \& {Jackson}}]{Demory2012}
{Demory}, B.-O., {Gillon}, M., {Seager}, S., {et~al.} 2012, \apjl, 751, L28

\bibitem[{{Demory} {et~al.}(2011){Demory}, {Gillon}, {Deming}, {Valencia},
  {Seager}, {Benneke}, {Lovis}, {Cubillos}, {Harrington}, {Stevenson}, {Mayor},
  {Pepe}, {Queloz}, {S{\'e}gransan}, \& {Udry}}]{Demory2011}
{Demory}, B.-O., {Gillon}, M., {Deming}, D., {et~al.} 2011, \aap, 533, A114

\bibitem[{{Demory} {et~al.}(2016{\natexlab{b}}){Demory}, {Gillon}, {de Wit},
  {Madhusudhan}, {Bolmont}, {Heng}, {Kataria}, {Lewis}, {Hu}, {Krick},
  {Stamenkovi{\'c}}, {Benneke}, {Kane}, \& {Queloz}}]{Demory2016b}
{Demory}, B.-O., {Gillon}, M., {de Wit}, J., {et~al.} 2016{\natexlab{b}}, \nat,
  532, 207

\bibitem[{{Donati}(2003)}]{espadons2003}
{Donati}, J.-F. 2003, in Astronomical Society of the Pacific Conference Series,
  Vol. 307, Solar Polarization, ed. J.~{Trujillo-Bueno} \& J.~{Sanchez
  Almeida}, 41

\bibitem[{{Donati} {et~al.}(1997){Donati}, {Semel}, {Carter}, {Rees}, \&
  {Collier Cameron}}]{Donati1997}
{Donati}, J.-F., {Semel}, M., {Carter}, B.~D., {Rees}, D.~E., \& {Collier
  Cameron}, A. 1997, \mnras, 291, 658

\bibitem[{{Dragomir} {et~al.}(2014){Dragomir}, {Matthews}, {Winn}, \&
  {Rowe}}]{Dragomir2014}
{Dragomir}, D., {Matthews}, J.~M., {Winn}, J.~N., \& {Rowe}, J.~F. 2014, in IAU
  Symposium, Vol. 293, Formation, Detection, and Characterization of Extrasolar
  Habitable Planets, ed. N.~{Haghighipour}, 52--57

\bibitem[{{Ehrenreich} {et~al.}(2012){Ehrenreich}, {Bourrier}, {Bonfils},
  {Lecavelier des Etangs}, {H{\'e}brard}, {Sing}, {Wheatley}, {Vidal-Madjar},
  {Delfosse}, {Udry}, {Forveille}, \& {Moutou}}]{Ehrenreich2012}
{Ehrenreich}, D., {Bourrier}, V., {Bonfils}, X., {et~al.} 2012, \aap, 547, A18

\bibitem[{{Fischer} {et~al.}(2008){Fischer}, {Marcy}, {Butler}, {Vogt},
  {Laughlin}, {Henry}, {Abouav}, {Peek}, {Wright}, {Johnson}, {McCarthy}, \&
  {Isaacson}}]{Fischer2008}
{Fischer}, D.~A., {Marcy}, G.~W., {Butler}, R.~P., {et~al.} 2008, \apj, 675,
  790

\bibitem[{{Gillon} {et~al.}(2012){Gillon}, {Demory}, {Benneke}, {Valencia},
  {Deming}, {Seager}, {Lovis}, {Mayor}, {Pepe}, {Queloz}, {S{\'e}gransan}, \&
  {Udry}}]{Gillon2012}
{Gillon}, M., {Demory}, B.-O., {Benneke}, B., {et~al.} 2012, \aap, 539, A28

\bibitem[{{Hu} \& {Seager}(2014)}]{Hu2014}
{Hu}, R., \& {Seager}, S. 2014, \apj, 784, 63

\bibitem[{{Ito} {et~al.}(2015){Ito}, {Ikoma}, {Kawahara}, {Nagahara},
  {Kawashima}, \& {Nakamoto}}]{Ito2015}
{Ito}, Y., {Ikoma}, M., {Kawahara}, H., {et~al.} 2015, \apj, 801, 144

\bibitem[{{Kite} {et~al.}(2016){Kite}, {Fegley}, {Schaefer}, \&
  {Gaidos}}]{Kite2016}
{Kite}, E.~S., {Fegley}, Jr., B., {Schaefer}, L., \& {Gaidos}, E. 2016, \apj,
  828, 80

\bibitem[{{Knutson} {et~al.}(2014){Knutson}, {Dragomir}, {Kreidberg},
  {Kempton}, {McCullough}, {Fortney}, {Bean}, {Gillon}, {Homeier}, \&
  {Howard}}]{Knutson2014}
{Knutson}, H.~A., {Dragomir}, D., {Kreidberg}, L., {et~al.} 2014, \apj, 794,
  155

\bibitem[{{Kreidberg} {et~al.}(2014){Kreidberg}, {Bean}, {D{\'e}sert},
  {Benneke}, {Deming}, {Stevenson}, {Seager}, {Berta-Thompson}, {Seifahrt}, \&
  {Homeier}}]{Kreidberg2014}
{Kreidberg}, L., {Bean}, J.~L., {D{\'e}sert}, J.-M., {et~al.} 2014, \nat, 505,
  69

\bibitem[{{Lammer} {et~al.}(2013){Lammer}, {Erkaev}, {Odert}, {Kislyakova},
  {Leitzinger}, \& {Khodachenko}}]{Lammer2013}
{Lammer}, H., {Erkaev}, N.~V., {Odert}, P., {et~al.} 2013, \mnras, 430, 1247

\bibitem[{{Lockwood} {et~al.}(2014){Lockwood}, {Johnson}, {Bender}, {Carr},
  {Barman}, {Richert}, \& {Blake}}]{Lockwood2014}
{Lockwood}, A.~C., {Johnson}, J.~A., {Bender}, C.~F., {et~al.} 2014, \apjl,
  783, L29

\bibitem[{{Madhusudhan} {et~al.}(2012){Madhusudhan}, {Lee}, \&
  {Mousis}}]{Madhusudhan2012}
{Madhusudhan}, N., {Lee}, K.~K.~M., \& {Mousis}, O. 2012, \apjl, 759, L40

\bibitem[{{Mandel} \& {Agol}(2002)}]{Mandel2002}
{Mandel}, K., \& {Agol}, E. 2002, \apjl, 580, L171

\bibitem[{{Martins} {et~al.}(2015){Martins}, {Santos}, {Figueira}, {Faria},
  {Montalto}, {Boisse}, {Ehrenreich}, {Lovis}, {Mayor}, {Melo}, {Pepe},
  {Sousa}, {Udry}, \& {Cunha}}]{Martins2015}
{Martins}, J.~H.~C., {Santos}, N.~C., {Figueira}, P., {et~al.} 2015, \aap, 576,
  A134

\bibitem[{{McArthur} {et~al.}(2004){McArthur}, {Endl}, {Cochran}, {Benedict},
  {Fischer}, {Marcy}, {Butler}, {Naef}, {Mayor}, {Queloz}, {Udry}, \&
  {Harrison}}]{McArthur2004}
{McArthur}, B.~E., {Endl}, M., {Cochran}, W.~D., {et~al.} 2004, \apjl, 614, L81

\bibitem[{{Nidever} {et~al.}(2002){Nidever}, {Marcy}, {Butler}, {Fischer}, \&
  {Vogt}}]{Nidever2002}
{Nidever}, D.~L., {Marcy}, G.~W., {Butler}, R.~P., {Fischer}, D.~A., \& {Vogt},
  S.~S. 2002, \apjs, 141, 503

\bibitem[{{Noguchi} {et~al.}(2002){Noguchi}, {Aoki}, {Kawanomoto}, {Ando},
  {Honda}, {Izumiura}, {Kambe}, {Okita}, {Sadakane}, {Sato}, {Tajitsu},
  {Takada-Hidai}, {Tanaka}, {Watanabe}, \& {Yoshida}}]{Noguchi2002}
{Noguchi}, K., {Aoki}, W., {Kawanomoto}, S., {et~al.} 2002, \pasj, 54, 855

\bibitem[{{Ridden-Harper} {et~al.}(2016){Ridden-Harper}, {Snellen}, {Keller},
  {de Kok}, {Di Gloria}, {Hoeijmakers}, {Brogi}, {Fridlund}, {Vermeersen}, \&
  {van Westrenen}}]{RiddenHarper2016}
{Ridden-Harper}, A.~R., {Snellen}, I.~A.~G., {Keller}, C.~U., {et~al.} 2016,
  \aap, 593, A129

\bibitem[{{Rodler} {et~al.}(2013{\natexlab{a}}){Rodler}, {K{\"u}rster}, \&
  {Barnes}}]{Rodler2013b}
{Rodler}, F., {K{\"u}rster}, M., \& {Barnes}, J.~R. 2013{\natexlab{a}}, \mnras,
  432, 1980

\bibitem[{{Rodler} {et~al.}(2013{\natexlab{b}}){Rodler}, {K{\"u}rster},
  {L{\'o}pez-Morales}, \& {Ribas}}]{Rodler2013a}
{Rodler}, F., {K{\"u}rster}, M., {L{\'o}pez-Morales}, M., \& {Ribas}, I.
  2013{\natexlab{b}}, Astronomische Nachrichten, 334, 188

\bibitem[{{Rodler} {et~al.}(2012){Rodler}, {Lopez-Morales}, \&
  {Ribas}}]{Rodler2012}
{Rodler}, F., {Lopez-Morales}, M., \& {Ribas}, I. 2012, \apjl, 753, L25

\bibitem[{{Rothman} {et~al.}(2010){Rothman}, {Gordon}, {Barber}, {Dothe},
  {Gamache}, {Goldman}, {Perevalov}, {Tashkun}, \& {Tennyson}}]{Rothman2010}
{Rothman}, L.~S., {Gordon}, I.~E., {Barber}, R.~J., {et~al.} 2010, \jqsrt, 111,
  2139

\bibitem[{{Sing} \& {L{\'o}pez-Morales}(2009)}]{Sing2009}
{Sing}, D.~K., \& {L{\'o}pez-Morales}, M. 2009, \aap, 493, L31

\bibitem[{{Sing} {et~al.}(2016){Sing}, {Fortney}, {Nikolov}, {Wakeford},
  {Kataria}, {Evans}, {Aigrain}, {Ballester}, {Burrows}, {Deming},
  {D{\'e}sert}, {Gibson}, {Henry}, {Huitson}, {Knutson}, {Lecavelier Des
  Etangs}, {Pont}, {Showman}, {Vidal-Madjar}, {Williamson}, \&
  {Wilson}}]{Sing2016}
{Sing}, D.~K., {Fortney}, J.~J., {Nikolov}, N., {et~al.} 2016, \nat, 529, 59

\bibitem[{{Snellen} {et~al.}(2014){Snellen}, {Brandl}, {de Kok}, {Brogi},
  {Birkby}, \& {Schwarz}}]{Snellen2014}
{Snellen}, I.~A.~G., {Brandl}, B.~R., {de Kok}, R.~J., {et~al.} 2014, \nat,
  509, 63

\bibitem[{{Snellen} {et~al.}(2010){Snellen}, {de Kok}, {de Mooij}, \&
  {Albrecht}}]{Snellen2010}
{Snellen}, I.~A.~G., {de Kok}, R.~J., {de Mooij}, E.~J.~W., \& {Albrecht}, S.
  2010, \nat, 465, 1049

\bibitem[{{Stevenson} {et~al.}(2014){Stevenson}, {D{\'e}sert}, {Line}, {Bean},
  {Fortney}, {Showman}, {Kataria}, {Kreidberg}, {McCullough}, {Henry},
  {Charbonneau}, {Burrows}, {Seager}, {Madhusudhan}, {Williamson}, \&
  {Homeier}}]{Stevenson2014}
{Stevenson}, K.~B., {D{\'e}sert}, J.-M., {Line}, M.~R., {et~al.} 2014, Science,
  346, 838

\bibitem[{{Swain} {et~al.}(2009){Swain}, {Tinetti}, {Vasisht}, {Deroo},
  {Griffith}, {Bouwman}, {Chen}, {Yung}, {Burrows}, {Brown}, {Matthews},
  {Rowe}, {Kuschnig}, \& {Angerhausen}}]{Swain2009}
{Swain}, M.~R., {Tinetti}, G., {Vasisht}, G., {et~al.} 2009, \apj, 704, 1616

\bibitem[{{Tajitsu} {et~al.}(2012){Tajitsu}, {Aoki}, \&
  {Yamamuro}}]{Tajitsu2012}
{Tajitsu}, A., {Aoki}, W., \& {Yamamuro}, T. 2012, \pasj, 64

\bibitem[{{Tamuz} {et~al.}(2005){Tamuz}, {Mazeh}, \& {Zucker}}]{Tamuz2005}
{Tamuz}, O., {Mazeh}, T., \& {Zucker}, S. 2005, \mnras, 356, 1466

\bibitem[{{Teske} {et~al.}(2013){Teske}, {Cunha}, {Schuler}, {Griffith}, \&
  {Smith}}]{Teske2013}
{Teske}, J.~K., {Cunha}, K., {Schuler}, S.~C., {Griffith}, C.~A., \& {Smith},
  V.~V. 2013, \apj, 778, 132

\bibitem[{{Tsiaras} {et~al.}(2016){Tsiaras}, {Rocchetto}, {Waldmann}, {Venot},
  {Varley}, {Morello}, {Damiano}, {Tinetti}, {Barton}, {Yurchenko}, \&
  {Tennyson}}]{Tsiaras2016}
{Tsiaras}, A., {Rocchetto}, M., {Waldmann}, I.~P., {et~al.} 2016, \apj, 820, 99

\bibitem[{{Valencia} {et~al.}(2010){Valencia}, {Ikoma}, {Guillot}, \&
  {Nettelmann}}]{Valencia2010}
{Valencia}, D., {Ikoma}, M., {Guillot}, T., \& {Nettelmann}, N. 2010, \aap,
  516, A20

\bibitem[{{Winn} {et~al.}(2011){Winn}, {Matthews}, {Dawson}, {Fabrycky},
  {Holman}, {Kallinger}, {Kuschnig}, {Sasselov}, {Dragomir}, {Guenther},
  {Moffat}, {Rowe}, {Rucinski}, \& {Weiss}}]{Winn2011}
{Winn}, J.~N., {Matthews}, J.~M., {Dawson}, R.~I., {et~al.} 2011, \apjl, 737,
  L18

\end{thebibliography}

%
\appendix
\begin{figure*}[h]
\begin{center}
%\scalebox{0.85}{\includegraphics{fig_reduct/fig_reduct_cfht1.eps}}
\scalebox{0.85}{\includegraphics{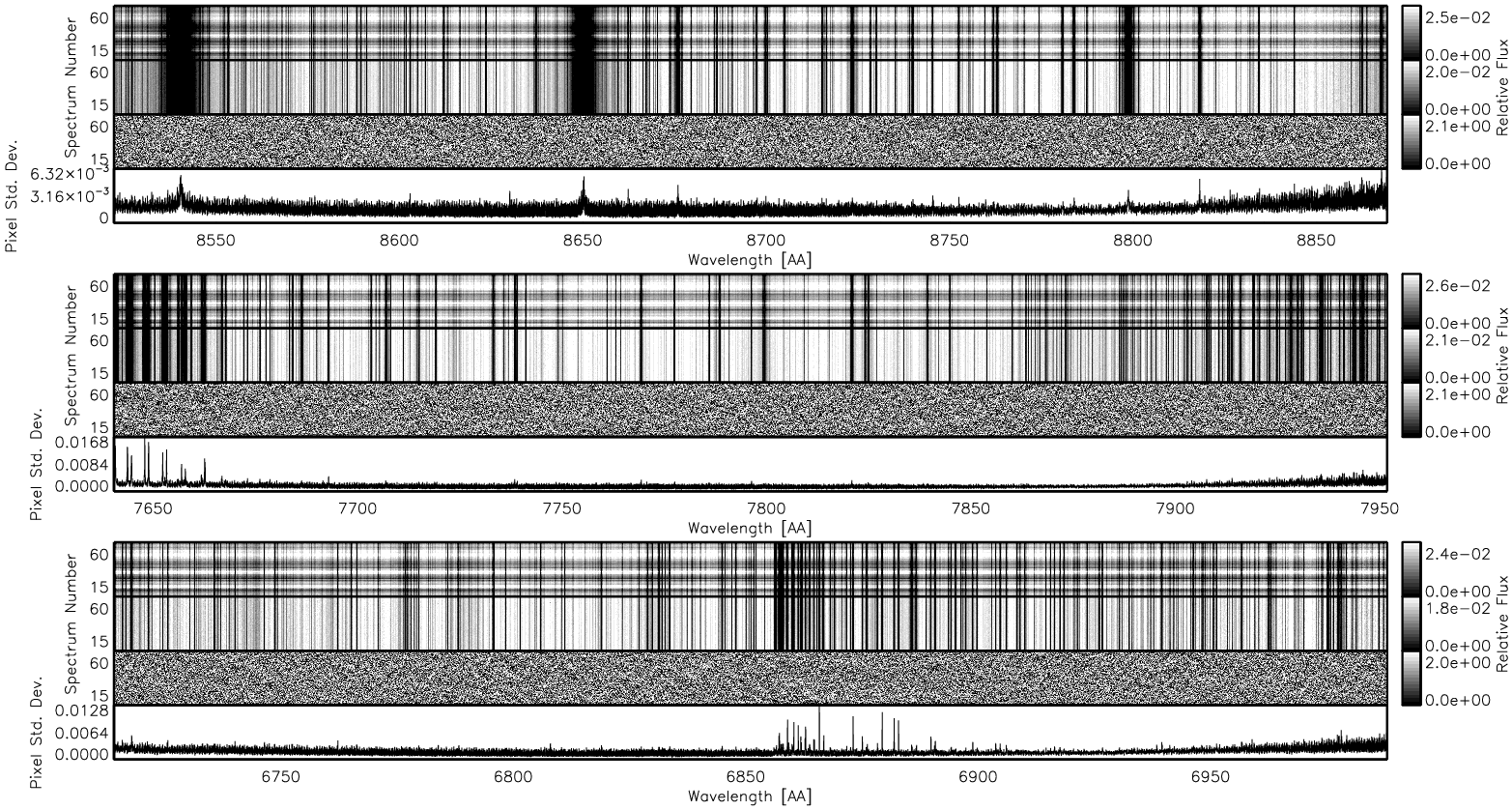}}
\end{center}
\caption{Top three panels: Stages of our reduction for a sample order of data from the first night of CFHT observations (N$_1$). First sub-panel from the top: The raw 1D spectra (see Section~\ref{sec:dr1}). Second sub-panel from the top: After normalization (see Section~\ref{sec:dr2}). Third sub-panel from the top: After stellar and telluric line removal (see Section~\ref{sec:dr2}). Fourth sub-panel from the top: The standard devation of each data point in the reduced spectra.}
\label{fig:reduct1}
\end{figure*}
\begin{figure*}[h]
\begin{center}
%\scalebox{0.85}{\includegraphics{fig_reduct/fig_reduct_cfht2.eps}}
\scalebox{0.85}{\includegraphics{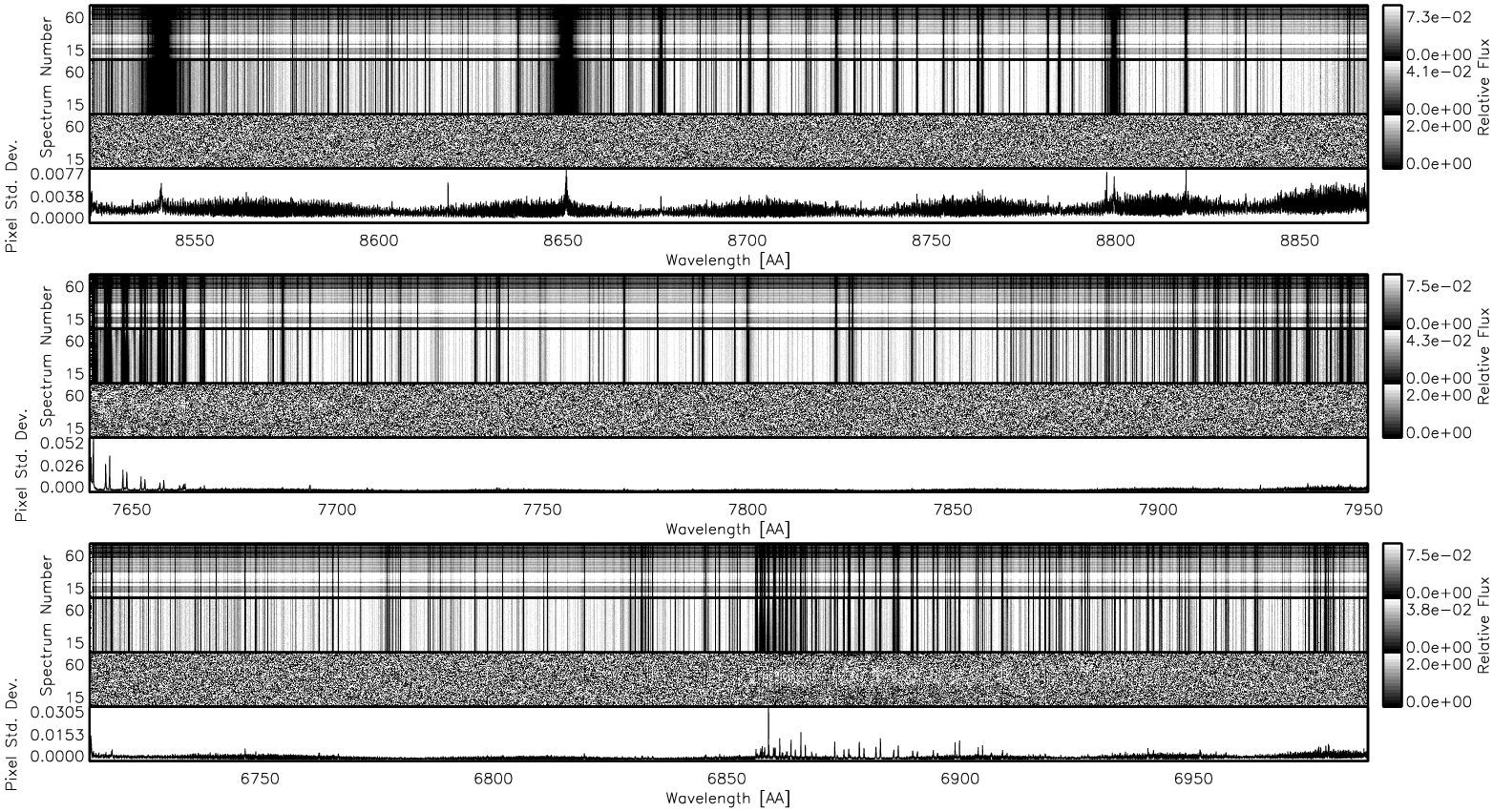}}
\end{center}
\caption{Each panel: Stages of our reduction for a sample order of data from the second night of CFHT observations (N$_2$). For a description of each sub-panel see Fig.~\ref{fig:reduct1}.}
\label{fig:reduct2}
\end{figure*}
\begin{figure*}[h]
\begin{center}
%\scalebox{0.85}{\includegraphics{fig_reduct/fig_reduct_subaru1.eps}}
\scalebox{0.85}{\includegraphics{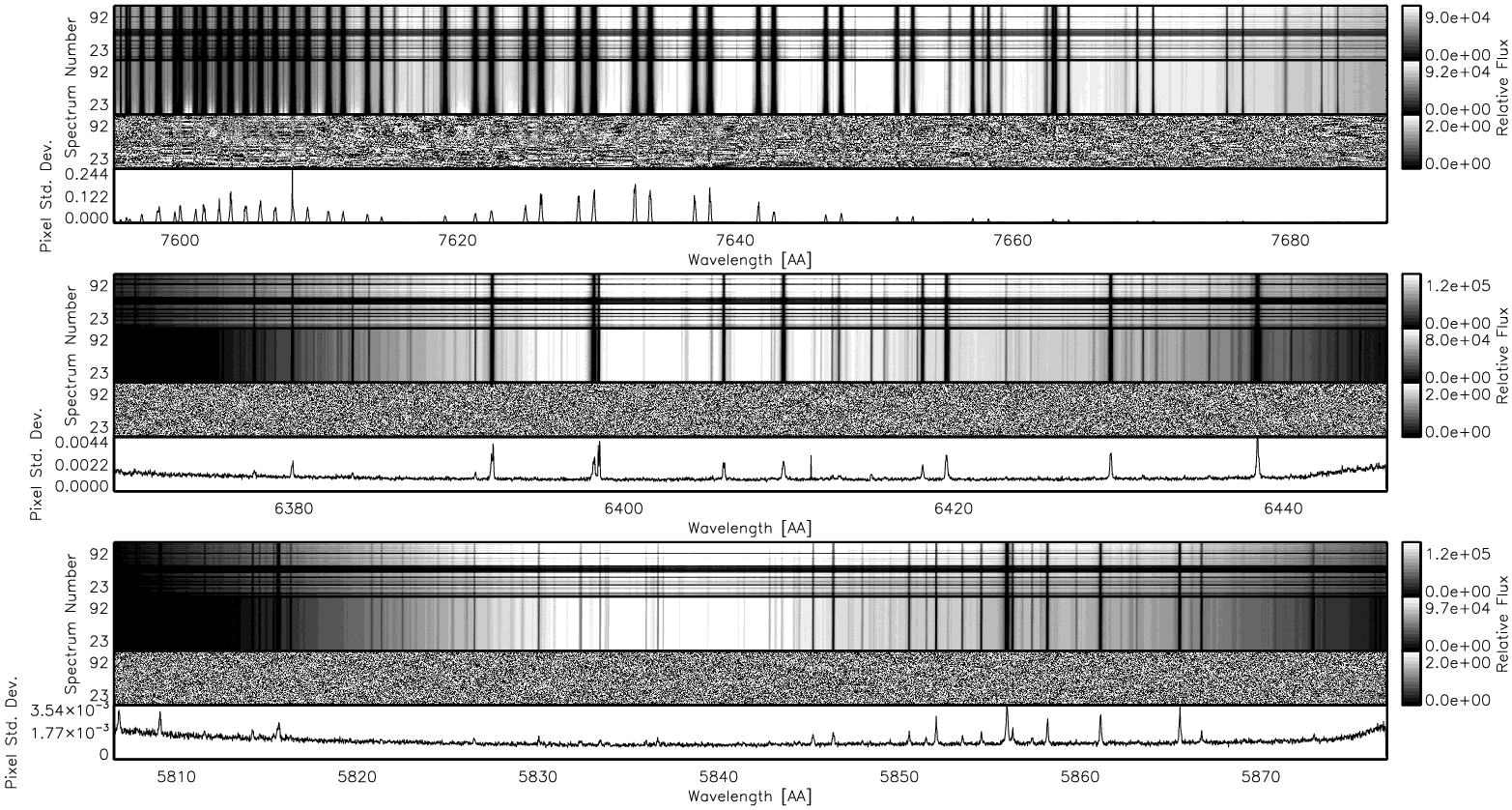}}
\end{center}
\caption{Each panel: Stages of our reduction for a sample order of data from the first night of Subaru observations (N$_3$). For a description of each sub-panel see Fig.~\ref{fig:reduct1}. }
\label{fig:reduct3}
\end{figure*}
\begin{figure*}[h]
\begin{center}
%\scalebox{0.85}{\includegraphics{fig_reduct/fig_reduct_subaru2.eps}}
\scalebox{0.85}{\includegraphics{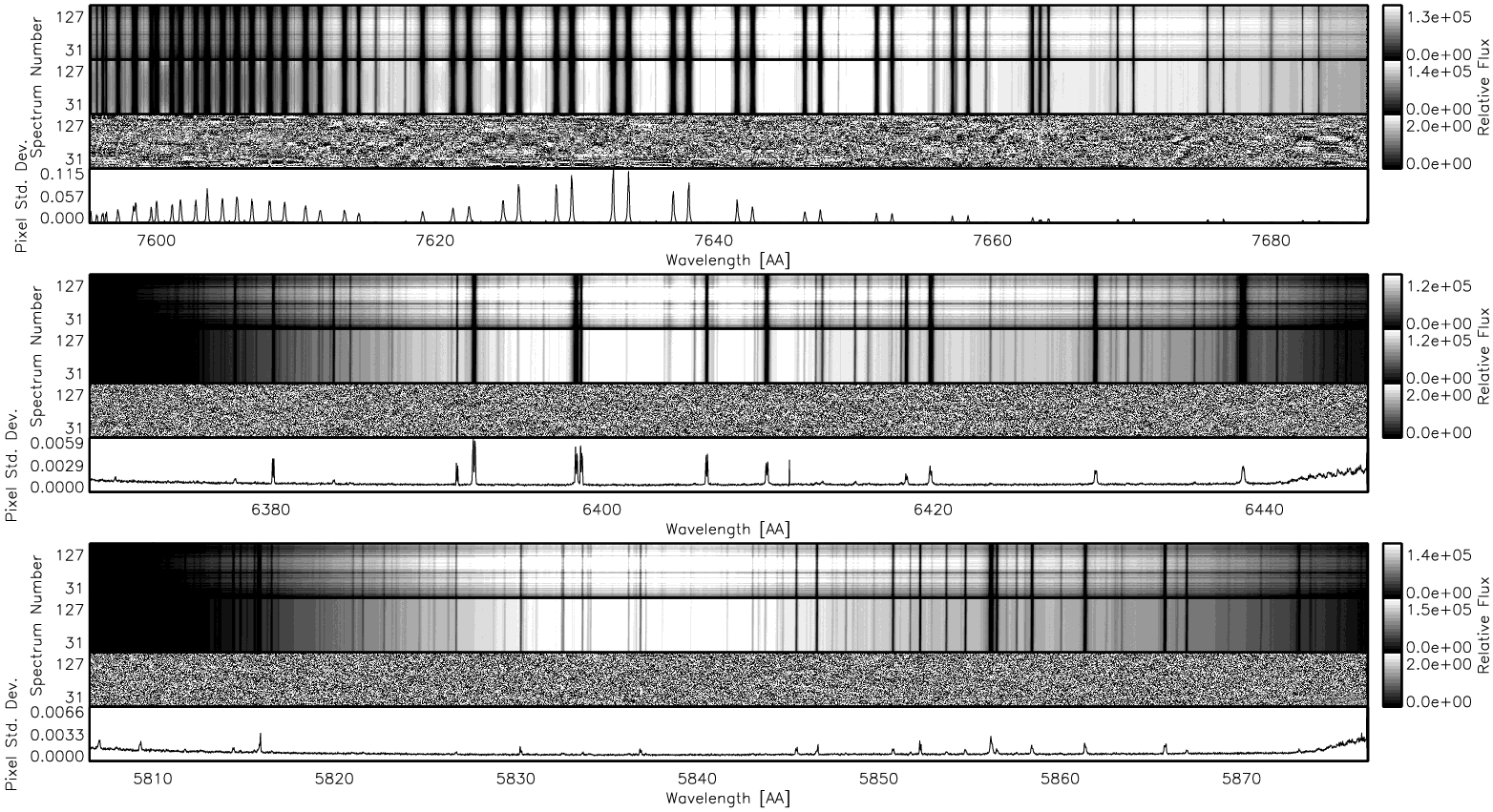}}
\end{center}
\caption{Each panel: Stages of our reduction for a sample order of data from the second night of Subaru observations (N$_4$). For a description of each sub-panel see Fig.~\ref{fig:reduct1}.}
\label{fig:reduct4}
\end{figure*}
\begin{figure*}[h]
\begin{center}
%\scalebox{0.85}{\includegraphics{fig_model/fig_models.eps}}
\scalebox{0.85}{\includegraphics{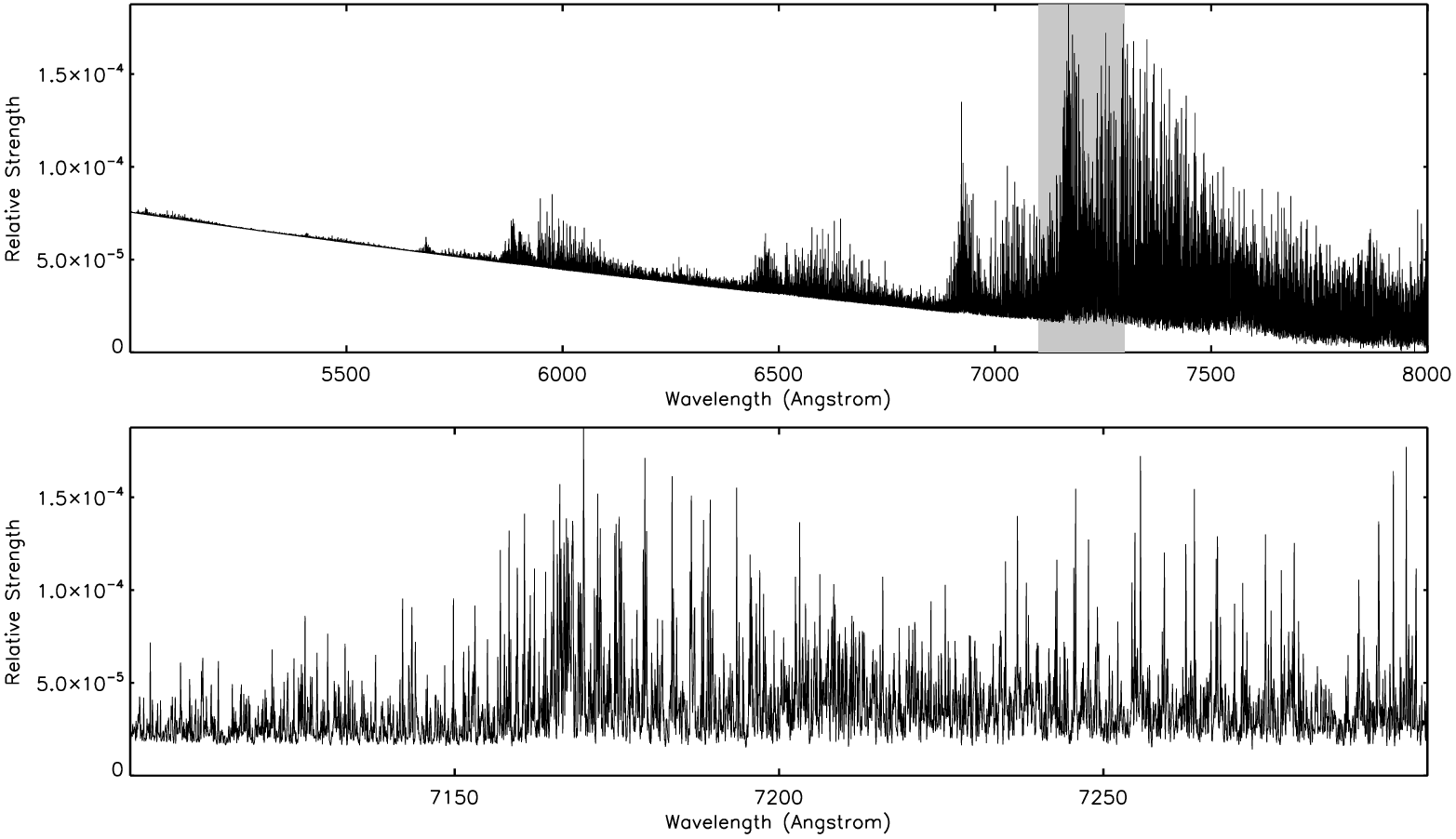}}
\end{center}
\caption{Top panel: An example of a model spectra for water absorption used in this analysis. The model correspondings to a water-rich atmosphere witha  mean molecular mass of 10 g/mol and a water volume mixing ratio of 10$\%$. Bottom panel: A close-up of the grey region of the spectra presented in the top panel.}
\label{fig:model}
\end{figure*}
\begin{figure*}[h]
\begin{center}
%\scalebox{0.85}{\includegraphics{fig_inject/fig_inject_before.eps}}
\scalebox{0.85}{\includegraphics{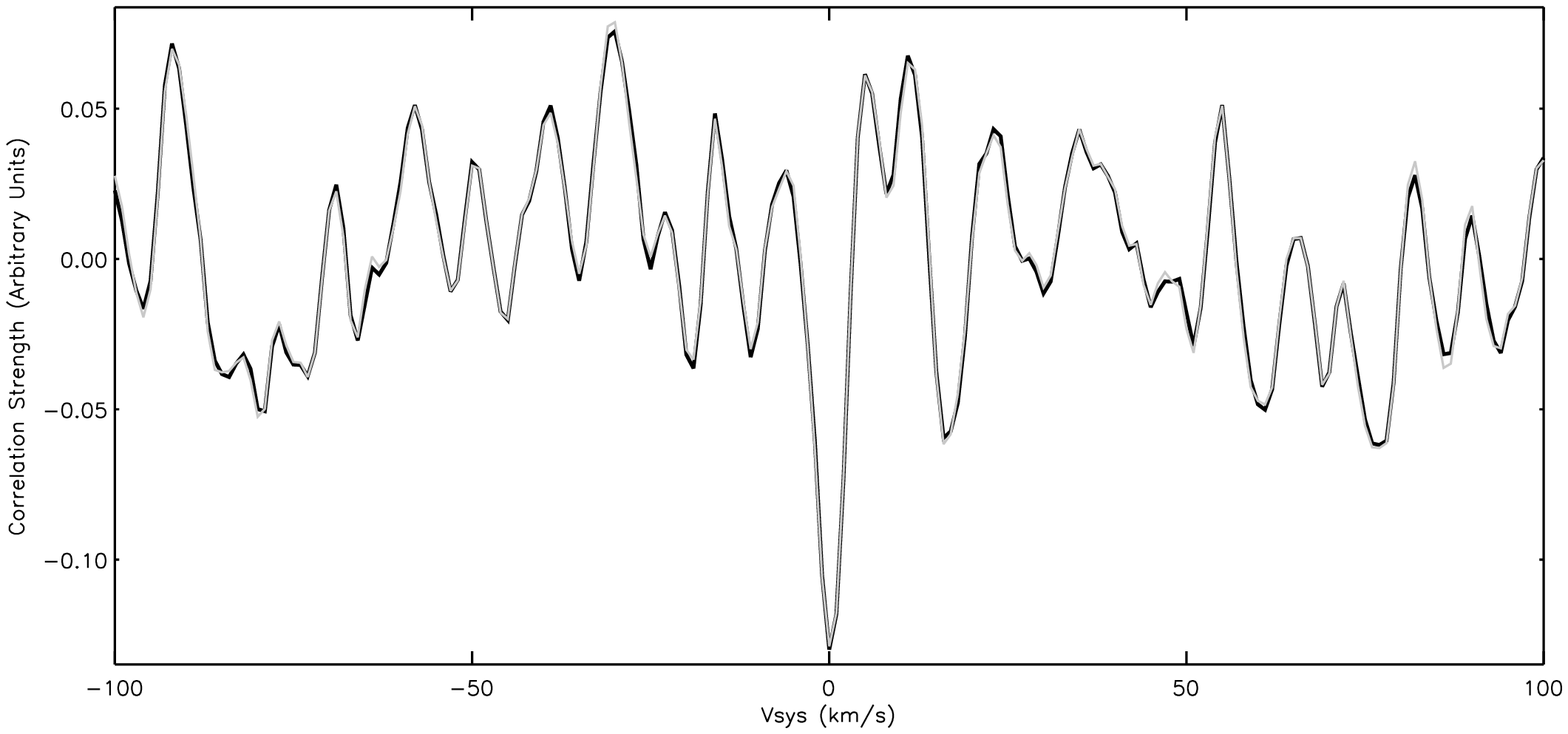}}
\end{center}
\caption{Correlation strength of the water-model injected data as a function of systemic velocity. The water model used has a mean molecular mass of 10 g/mol and a water VMR of 10\%. In black is the correlation of data with the model injected before interpolating to a common wavelength grid, while grey is the correlation with the model injected after the interpolation.}
\label{fig:inject_before}
\end{figure*}
\end{document}